\documentclass[a4paper,fleqn,useAMS,usenatbib]{mnras}

\usepackage{graphicx}	
\usepackage{amsmath}	
\usepackage{amssymb}	
\usepackage{multicol}        
\usepackage{bm}		
\usepackage{pdflscape}	
\usepackage{xspace}
\usepackage[final]{pdfpages}
\usepackage{layouts}

\usepackage{ulem}

\newcommand{\ec}{$\eta$~Car\xspace}
\newcommand{\brite}{{\it BRITE}-Constellation\xspace}

\title[Pulsations in $\eta$ Carinae]{{\it BRITE}-Constellation reveals evidence for pulsations in the enigmatic binary $\eta$ Carinae}

\author[Richardson et al.]{Noel D.~Richardson$^1$\thanks{E-mail:noel.richardson@utoledo.edu}
Herbert Pablo$^{2}$,
Christiaan Sterken$^{3}$,
Andrzej Pigulski$^{4}$,
 \newauthor Gloria Koenigsberger$^{5}$,
Anthony F. J. Moffat$^{6}$,
Thomas I. Madura$^{7}$,
 \newauthor Kenji Hamaguchi$^{8,9}$,
Michael F. Corcoran$^{8,10}$,
Augusto Damineli$^{11}$,
 \newauthor Theodore R. Gull$^{12}$,
D. John Hillier$^{13}$,
Gerd Weigelt$^{14}$,
Gerald Handler$^{15}$,
 \newauthor Adam Popowicz$^{16}$,
Gregg A. Wade$^{17}$,
Werner W. Weiss$^{18}$,
and Konstanze Zwintz$^{19}$\\
$^{1}$ Ritter Observatory, Department of Physics and Astronomy, The University of Toledo, Toledo, OH 43606-3390, USA\\
$^{2}$ AAVSO Headquarters, 49 Bay State Rd., Cambridge, MA 02138 \\
$^{3}$ Department of Physics, Vrije Universiteit Brussel, Pleinlaan 2, 1050 Brussels, Belgium\\
$^{4}$ Instytut Astronomiczny, Uniwersytet Wroclawski, Kopernika 11, 51-622, Wroclaw, Poland \\
$^{5}$ Instituto de Ciencias F\'isicas, Universidad Nacional Aut\'onoma de M\'exico, Ave. Universidad S/N, Cuernavaca, 62210, Morelos,\\ M\'exico\\
$^{6}$ D\'epartement de physique and Centre de Recherche en Astrophysique du Qu\'ebec (CRAQ), Universit\'e de Montr\'eal, C.P. 6128,\\  Succ.~Centre-Ville, Montr\'eal, Qu\'ebec, H3C 3J7, Canada\\
$^{7}$ San Jos\'e State University, Department of Physics and Astronomy, One Washington Square, San Jos\'e, CA 95192-0106, USA \\
$^{8}$ CRESST II and X-ray Astrophysics Laboratory NASA/GSFC, Greenbelt, MD 20771, USA\\
$^{9}$ Department of Physics, University of Maryland, Baltimore County, 1000 Hilltop Circle, Baltimore, MD 21250, USA \\
$^{10}$ Institute for Astrophysics and Computational Sciences, Department of Physics, The Catholic University of America,\\ Washington, DC 20064, USA\\
$^{11}$ Instituto de Astronomia, Geof'sica e Ci\^encias Atmosf\'ericas, Universidade de S\~{a}o Paulo, Rua do Mat\~{a}o 1226, Cidade\\ Universit\'aria, S\~{a}o Paulo, 05508-900, Brazil\\
$^{12}$ Code 667, NASA Goddard Space Flight Center, Greenbelt, MD 20771, USA \\
$^{13}$ Department of Physics and Astronomy \& Pittsburgh Particle Physics, Astrophysics, and Cosmology Center (PITT PACC),\\ University of Pittsburgh, Pittsburgh, PA 15260, USA\\
$^{14}$ Max Planck Institute for Radio Astronomy, Auf dem H\"{u}gel 69, D-53121 Bonn, Germany\\
$^{15}$ Nicolaus Copernicus Astronomical Center, ul. Bartycka 18, 00-716, Warsaw, Poland \\
$^{16}$ Institute of Automatic Control, Silesian University of Technology, Akademicka 16, 44-100 Gliwice, Poland \\
$^{17}$ Department of Physics and Space Science, Royal Military College of Canada, PO Box 17000 Kingston, ON K7K 7B4, Canada \\
$^{18}$ Institute for Astrophysics, University of Vienna, Tuerkenschanzstrasse 17, 1180 Vienna, Austria \\
$^{19}$ Universit{\"a}t Innsbruck, Institut fur Astro- und Teilchenphysik Technikerstrasse 25/8, A-6020 Innsbruck }

\begin{document}

\date{}

\pagerange{\pageref{firstpage}--\pageref{lastpage}} \pubyear{2017}

\maketitle

\label{firstpage}

\begin{abstract}
\ec is a massive, eccentric binary with a rich observational history. We obtained the first high-cadence, high-precision light curves with the {\it BRITE}-Constellation nanosatellites over 6 months in 2016 and 6 months in 2017. The light curve is contaminated by several sources including the Homunculus nebula and neighboring stars, including the eclipsing binary CPD\,$-$59$^\circ$2628. However, we found two coherent oscillations in the light curve. These may represent pulsations that are not yet understood but we postulate that they are related to tidally excited oscillations of $\eta$ Car's primary star, and would be similar to those detected in lower-mass eccentric binaries. In particular, one frequency was previously detected by van Genderen et al.~and Sterken et al.~through the time period of 1974 to 1995 through timing measurements of photometric maxima. Thus, this frequency seems to have been detected for nearly four decades, indicating that it has been stable in frequency over this time span. These pulsations could help provide the first direct constraints on the fundamental parameters of the primary star if confirmed and refined with future observations.
\end{abstract}


\section{Introduction}

Most massive stars are found in binary systems, with a majority of these binaries experiencing interactions during the stellar lifetimes \citep{2012Sci...337..444S}. One of the most enigmatic binaries in the Galaxy is $\eta$ Carinae, which has a massive luminous blue variable (LBV) primary star. The secondary star has eluded observers since the discovery of the 5.54 year periodicity by \citet{1996ApJ...460L..49D} but some properties have been inferred through X-ray analyses of the colliding winds \citep[e.g., ][]{2001ApJ...547.1034C, 2017ApJ...838...45C}. The rapid changes in the spectrum that occur every 5.54 years are caused by energetic phenomena that occur at the periastron passage of a highly eccentric system \citep[$e\sim 0.9$; see][]{2008MNRAS.386.2330D, 2016ApJ...819..131T}. Some recent observations indicate that there may be long-term changes present within the system \citep[e.g., ][]{2010ApJ...717L..22M, 2012ApJ...751...73M}.

\ec is one of the most studied objects in the sky. Photometric observations can be traced to before the Great Eruption that occurred in 1843 \citep[e.g.,][]{1994PASP..106.1025H,2011MNRAS.415.2009S}, when the system expelled many solar masses of material to create the bipolar Homunculus nebula that surrounds the system today. After dust began to form and the Great Eruption ended, the system faded to below naked eye visibility except for a brief brightening due to a second, smaller eruption in the 1890s. Since then, the system has been slowly growing in brightness as the dust dissipates, and is currently at a $V$-band magnitude near 4.2. There are excursions in the brightness observed when the stars are near periastron, as reported by \citet{2004MNRAS.352..447W} and \citet{2010NewA...15..108F}. 

This photometric variability near periastron could have several causes. \citet{2010RMxAC..38...52M} postulated that this signature was caused by the wind-wind collision region carving a hole in the primary's dense wind pseudo-photosphere during periastron passage, allowing radiation from the inner layers to escape more easily. However, brightening due to tidal shear energy dissipation around periastron passage has been postulated to cause periastron events by \citet{2011A&A...528A..48M}.  Indeed, the eclipse-like event in the light curve occuring near periastron \citep[see][]{2010NewA...15..108F} also bears a striking similarity to the heartbeat phenomenon discovered in eccentric binaries with the {\it Kepler} Space Telescope \citep{2012ApJ...753...86T} and which is caused by tidal effects.

The heartbeat phenomenon occurs in close eccentric binaries where the changes in the gravitational potential introduce strong perturbations in the tidal amplitude and configuration. While {\it Kepler} only observed this phenomenon in stars with masses up to about twice the mass of the sun, it has recently been discovered in the eccentric OB system $\iota$ Orionis ($P=29.13376$ d, $e=0.764$, $M_1 = 23.2 M_\odot$) by \citet{PABLO} with the \brite nanosatellites, and was predicted to be present by \citet{1999RMxAA..35..157M}. Other massive stars may also show this phenomenon, and with the large datasets being collected by \brite and the {\it K2} missions, more will likely be discovered in the near future. 

In addition to the tidal distortion of the stars near periastron, heartbeat systems exhibit pulsational behavior that is driven by the binary orbit and the changing tidal forces during the orbit. These pulsations occur at resonant harmonics of the orbital period \citep{1995ApJ...449..294K, 2011ApJS..197....4W}. Analysis of these tidally-induced pulsations can help constrain masses, radii, and fundamental parameters even without precise radial velocities or photometric eclipses. After \citet{1997NewA....2..107D} proposed a binary orbit as an explanation for spectroscopic events in \ec, \citet{1997NewA....2..387D} noted that the eccentricity must be very high to fit an orbit. In such a case, \citet{1997NewA....2..387D} predicted that tidal forces would be important if $e \gtrsim 0.85$, which was also discussed again recently by \citet{2017RNAAS...1....6D}.

In this paper, we report on observations of \ec obtained with \brite during the years 2016 and 2017 (Section 2). Section 3 shows how these observations are found to exhibit coherent variations at the millimag-level across the two six-month duration time-series. \ec lies in a crowded field, and we discuss the contaminants in the light curve in Section 4. We discuss the implications in Section 5, and the need for future work, both observationally and theoretically, in Section 6.

\section{BRITE Photometry}

Precision time-series photometry was collected for this project with {\it BRITE} ({\it BRIght Target Explorer}) - Constellation. This is a fleet of five nanosatellites described in detail by \citet{2014PASP..126..573W} and \citet{2016PASP..128l5001P}. The telescopes have 3 cm apertures, with two satellites recording blue images (3900--4600\AA) and three satellites recording red images (5500--7000\AA). The filters are similar to the Geneva $B$ and Sloan $r$ filters. The satellites' field of view is large, $20^\circ \times24^\circ$. Data are downloaded for over 30 stars simultaneously. The orbital periods of the fleet range between 97 and 101 minutes. The satellite images have large (28\arcsec) pixels on the CCDs, with a FWHM between 5--8 pixels, meaning that the light of \ec blurs into the light of the bipolar Homunculus nebula (with a size of 18\arcsec) during the observations. This dilution will dilute the strength of the any stellar variability with the contamination from the surrounding Homunculus nebula. The amount of this dilution remains unknown as no high-resolution imaging capable of separating the flux of the central source and the Homunculus nebula has been obtained recently. Further, the actual amplitude could be diluted intrinsically by the strong stellar wind. 

A total of 57127 data points were recorded with the {\it BRITE-Toronto} (BTr; red filter) satellite between 2016 February 10 and 2016 July 23 \citep[phase 13.28--13.36 according to the ephemeris of][]{2016ApJ...819..131T}. The measurements were made with four- or five-second exposures for periods of 5--10 minutes per orbit when the field was visible. The chopping mode that was utilized allows us to perform photometry through a difference between successive frames with the star moving to different positions in the sub-raster. The resulting difference image yields a positive and negative stellar PSF with bad pixels and sky background subtracted. This technique increases data quality and precision \citep{2016SPIE.9904E..1RP, 2017arXiv170509712P}. Precision was further increased by creating a mean measurement for each satellite orbit of 98.24 minutes, resulting in an average 1.2 mmag r.m.s.~precision for the differential photometry after removing correlations within the data, as described by \citet{2016PASP..128l5001P}. This resulted in a total of 1631 photometric points. 

Similar observations of \ec were obtained in 2017 with both {\it BRITE-Heweliusz} (BHr) and {\it UniBRITE} (UBr), which also operate with a red filter identical to that on {\it BRITE-Toronto}. The resulting light curve from {\it UniBRITE} had 884 points binned on the satellite orbits taken between 2017 January 11 and 2017 May 4. 1056 points were obtained with {\it BRITE-Heweliusz} between 2017 March 30 and 2017 July 1. The reductions show identical variability in the time frame between 2017 March 30 and 2017 May 4 when both of these satellites observed the source. The second year of observations show typical errors of 2.2--2.5 mmag per orbit-binned data point, owing to the shorter exposures being necessary with these two satellites.

There are many ways to obtain a reduced light curve from \brite, as many different correlations can exist within the measurements. This is discussed by \citet{2016PASP..128l5001P}. Given the unusual nature of the LBV primary, we elected to do the minimal amount of detrending within the data, to better retain longer-period intrinsic variability inherent to the LBV-like object. As no ground-based long-term photometric monitoring was collected in parallel to these observations, we cannot check for consistency and accuracy in the final light curve as long-term changes are present in LBVs. This may result in derived peaks from the Fourier transformations having inaccurate amplitudes.

\begin{figure*}
\includegraphics[angle=90,width=3.3in]{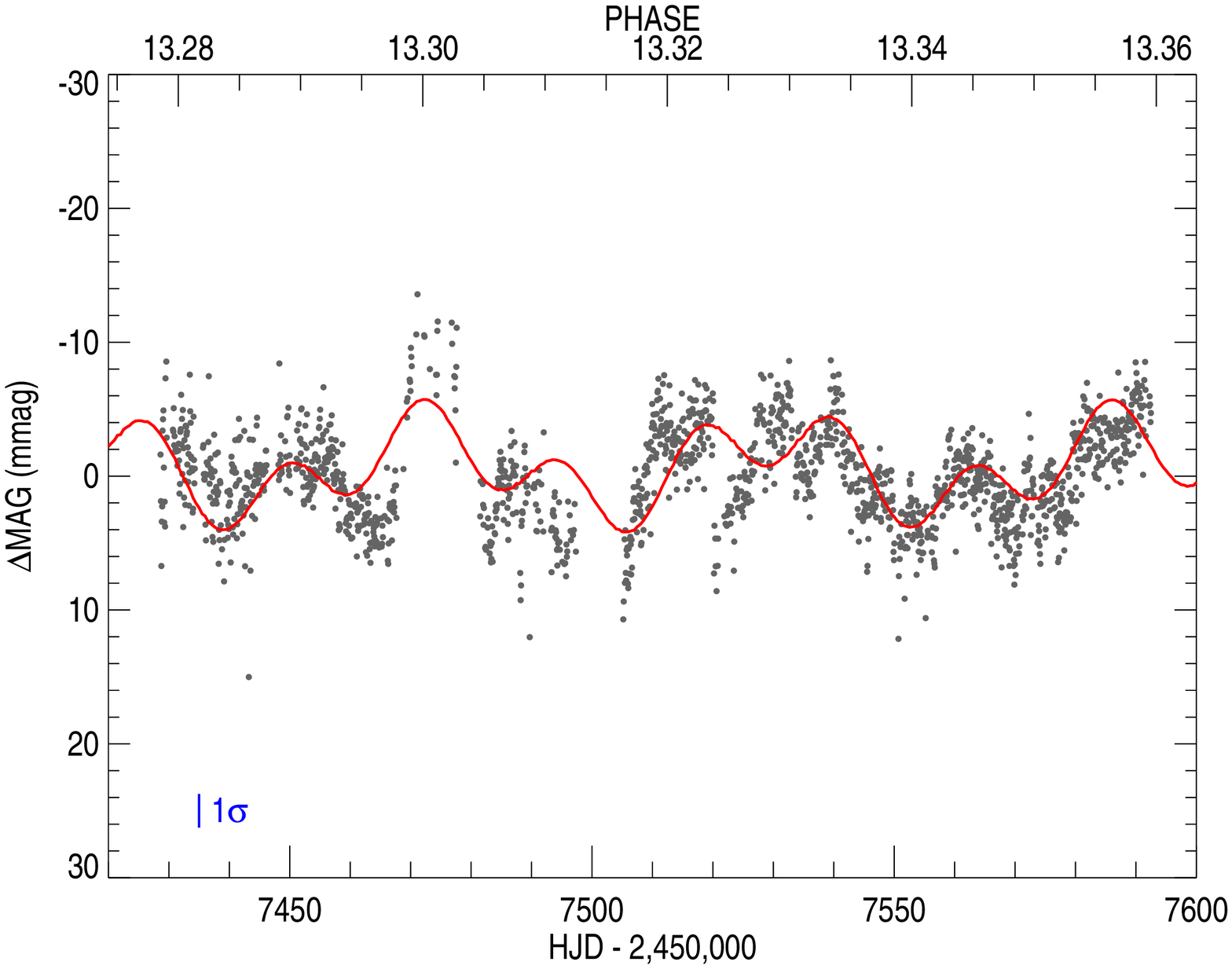}
\includegraphics[angle=90,width=3.3in]{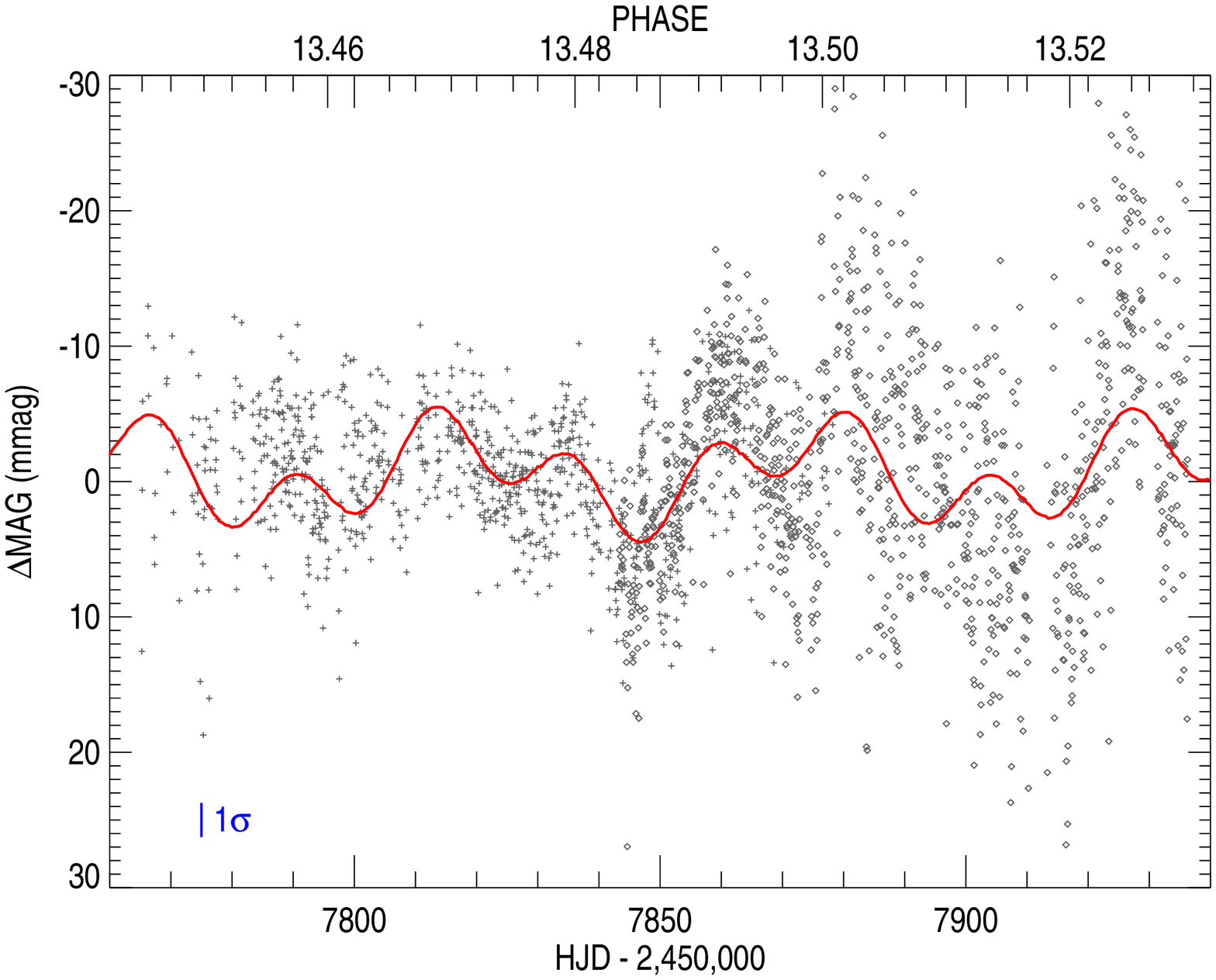}

\caption{The light curves observed by {\it BRITE} in 2016 (left) and 2017 (right). The data points are solid circles ($\bullet$; BTr), plus signs ($+$; UBr), and open diamonds ($\Diamond$; BHr) with the two-frequency fit shown in red. Each panel shows 180 d of time on the abscissa, with nearly 180 d between the two panels. A typical 2 mmag error bar (2$\sigma$) are shown in the lower left part of the panels.}
\label{lc}
\end{figure*}

\section{Variability found through Fourier Transformation}

\begin{figure*}
\includegraphics[angle=90,width=3.3in]{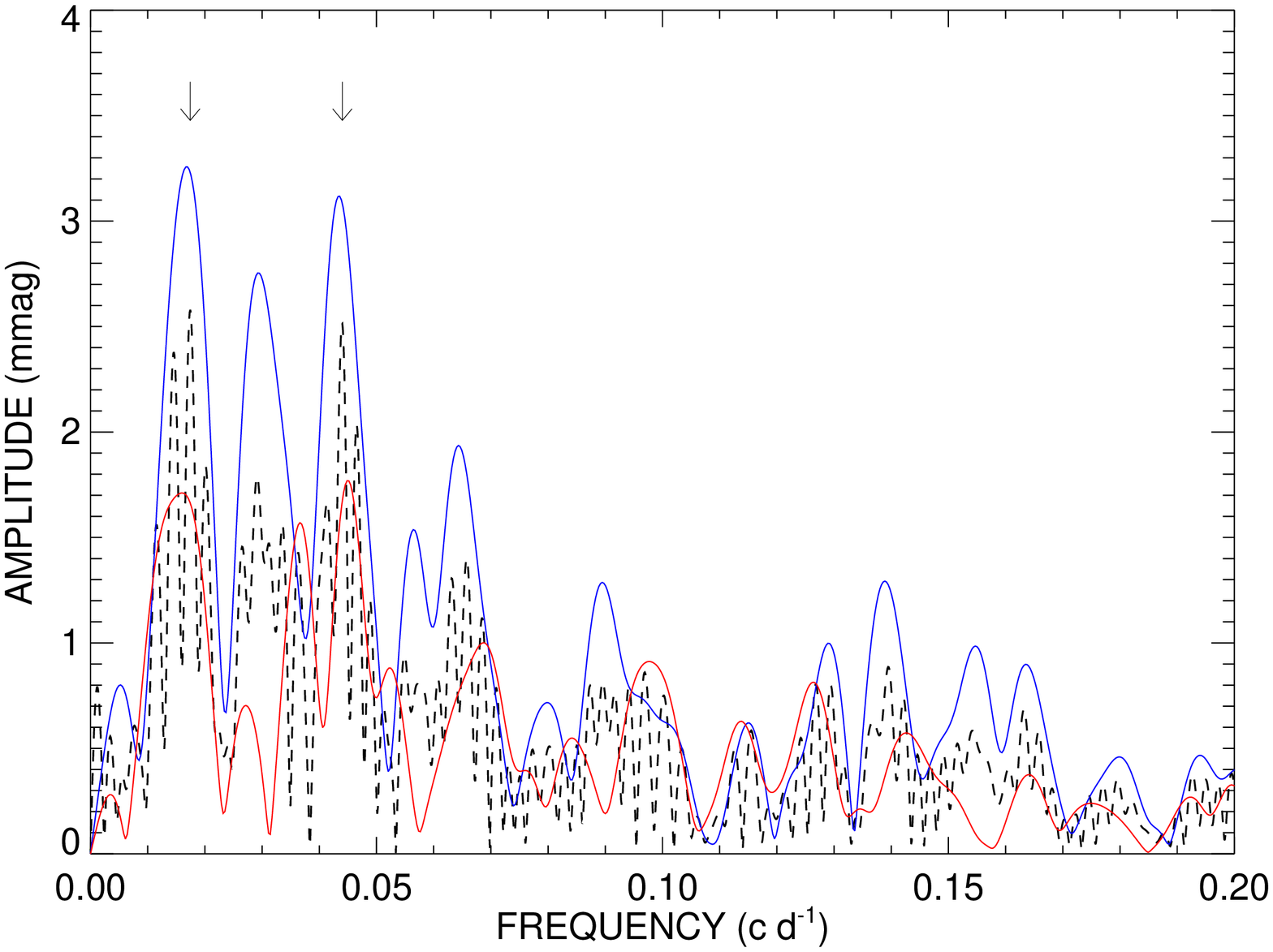}
\includegraphics[angle=90,width=3.57in]{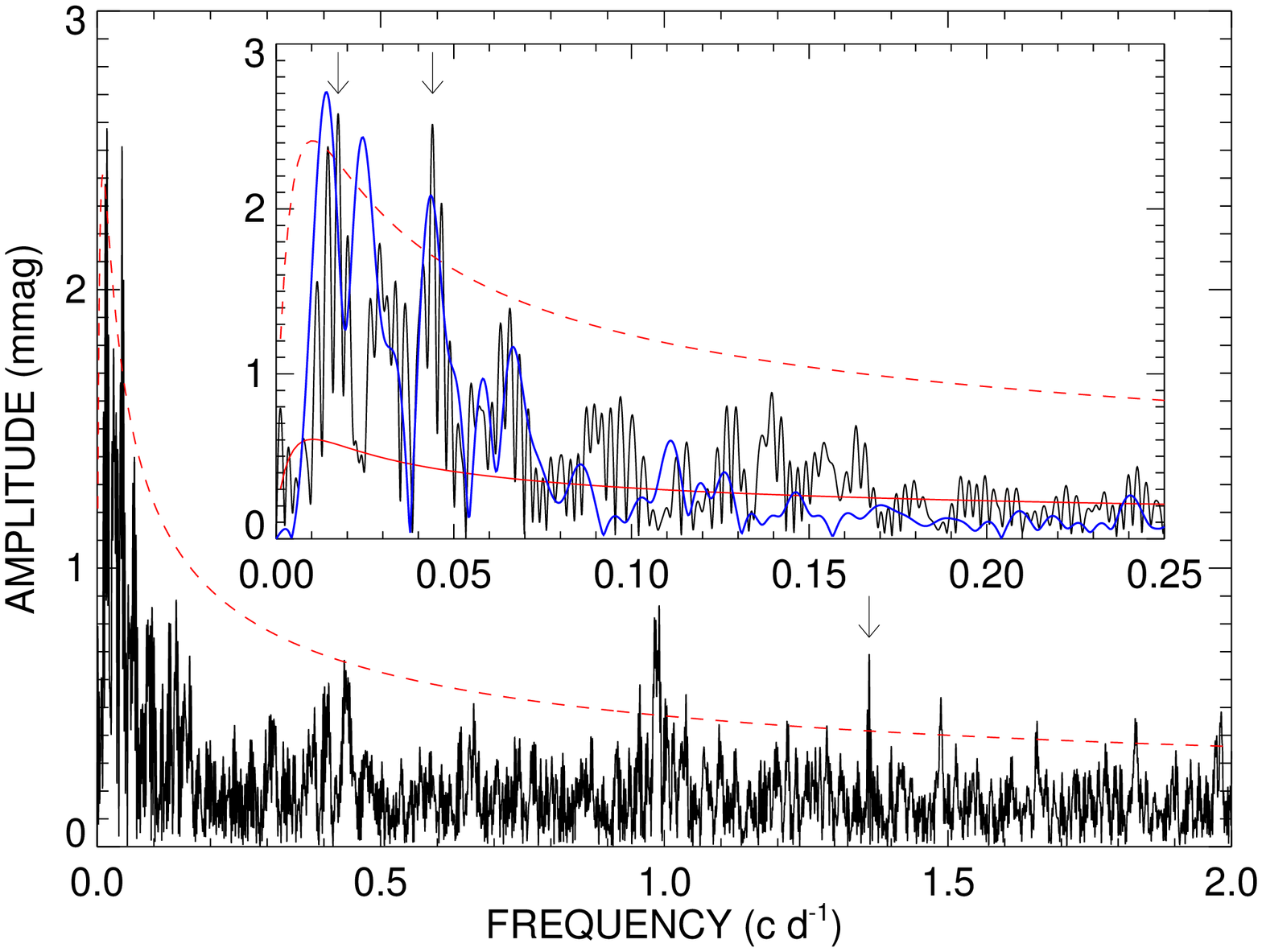}
\caption{Fourier amplitude spectrum of the photometric data. On the left, we show the Fourier spectrum of the 2016 data (blue solid line), the 2017 data (red solid line), and the combined dataset (black dashed line). Unsurprisingly, the combined dataset provides similar peaks to the peaks from the individual seasons. In the right panel, we show the Fourier amplitude spectrum of the combined data set, and indicate the frequency that represents a harmonic of the orbital period of CPD\,$-$59$^\circ$2628. In the inset plot, we show the noise floor in red, along with a 4$\sigma$ detection threshold in a dashed red line. The two periods from Table \ref{table} are indicated. In blue, we show a Fourier amplitude spectrum of the TIDES model discussed in the text.}
\label{FT}
\end{figure*}

The \brite light curves are plotted in Fig.~\ref{lc} as a function of both time and orbital phase. The phase was calculated using the ephemeris of \citet{2016ApJ...819..131T}, which represents the predicted periastron passage from the binary models. The leading number 13 is in reference to the first-observed ``low-ionization" event, which would be relative to the periastron passage that was observed in 1942 and reported by \citet{1953ApJ...118..234G}. The variability is clearly seen to be at a higher amplitude than the errors of the individual points, so we began our analysis by using the {\tt Period04} software \citep{2005CoAst.146...53L} to calculate the amplitude of the Fourier transform of the data, shown in Fig. \ref{FT}. Each frequency is of the form
$ A_i \sin (2\pi (\omega t +\phi_i)) ,$
where $A_i$ is the amplitude, $\omega$ is the angular frequency, and $\phi$ is the offset phase. These frequencies can be improved upon after fitting other peaks in the Fourier transform. Our values of $\phi$ are based on the zero point of the HJD times.
This approach allows the user to fit the selected peak in the Fourier amplitude spectrum and remove it from the dataset. Through an iterative process, we then infer multiple frequencies from the light curve, their amplitudes, and phase information. We continued this process of successive pre-whitening until the remaining Fourier peaks were near the noise floor and were not statistically significant. There is a small peak at 1 cycle d$^{-1}$ which is caused by the daily changes in terrestrial illumination as the satellites orbit the Earth.

\begin{table*}
\begin{minipage}{170mm}
\centering
\caption{Derived Frequencies \label{table}}
\begin{tabular}{l  c  c c c c c}
\hline \hline
		&	Frequency &	Period &	Amplitude	&	Phase	&	S/N	&	Potential  \\ 
		&	(c d$^{-1}$) & 	(d)	&	(mmag)	&			&		&	Harmonics \\ \hline

$f_1$	&	0.0174(5)	&	58.8		&	2.64	&	0.551	&	4.52	&	34--36	\\
$f_2$	&	0.0440(5)	&	22.7		&	2.47	&	0.042	&	5.86	&	87--91	\\
\hline
\end{tabular}
\end{minipage}
\end{table*}

Also shown in Fig. \ref{FT} is the noise floor for the Fourier amplitude spectrum, similar to that described by \citet{PABLO}. The noise floor represents the level at which no reliable signal is fit. It also neglects any contribution from the actual periods that are present in the data, so may be over-estimated at the low-frequency end. To calculate the noise present in the Fourier transform, we used a procedure outlined by \citet{PABLO} and based upon that of \citet{2002A&A...390.1119G}. We use Gaussian statistics to define a false alarm probability (FAP) in a series of $N$ frequency bins, where a peak would be $m\times$ above the noise. Then, 
$$ FAP_N (m) = 1 - (1-e^{-m})^{pN}.$$
Here $p$ is set equal to 2.8, and represents an empirical term necessary because we oversample the Fourier transform. In order to calculate a significance threshold, we choose a value of $FAP=0.05\%$. This is shown as a solid red line in the right, inlaid panel of Fig.~\ref{FT}. We also plot a dashed red line, which is four times this level of significance in the right panel of Fig.~\ref{FT}. 
The two derived significant frequencies are listed in Table \ref{table}, along with their corresponding periods and amplitudes. The signal-to-noise of the peaks are about 4$\times$ the noise floor. These values of significance may be higher as the noise floor will be over-estimated for these low frequencies as it was calculated prior to the removal of the two frequencies. With additional data taken by \brite in the future, the significance of these peaks should increase.

\section{Contamination within the light curve}

\ec is a member of the young cluster Trumpler\,16 and lies in a crowded region of the Carina nebula. While most of the neighboring stars or objects do not emit any detectable light in the optical, the Homunculus nebula and seven stars or stellar systems are all brighter than $V=10$. All of these stars have a spectral type listed in SIMBAD of O.

The causes of  O-star photometric variability are many and not always simple to identify. \citet{2014MNRAS.441..910R} found that the mid-O star $\xi$ Per exhibited photometric behavior attributed to spots on the stellar surface that may drive the discrete absorption component phenomenon often observed in O stars \citep[e.g., ][]{1996A&AS..116..257K}. Similar phenomena were recently confirmed with both photometry and spectroscopy for the early-O supergiant $\zeta$ Puppis \citep{2017arXiv171008414R}. The late-O supergiant $\zeta$ Orionis \citep{2017A&A...602A..91B} shows both a rotational modulation from its magnetic field and a secondary component that varies on a longer timescale and may be associated with the stellar wind.

The comparison stars used in ground-based photometric monitoring of \ec \citep{1995A&A...304..415V, 1996A&AS..116....9S, 1999A&A...343..847V} included HDE 303308 (O3V), CPD\,$-$59$^\circ$2632 (B1V), and CPD\,$-$59$^\circ$2627 (O8.5V). While all these stars have a fairly non-variable light curve as observed from the ground, the spectral types imply that they are likely to show millimag-level variations due to intrinsic variability such as that reported by \citet{2017A&A...602A..91B} or \citet{2014MNRAS.441..910R}. These stars are likely included in our photometric data as they lie within 3-4\arcmin of $\eta$ Car. However, it is difficult to see directly if their flux lies within the PSF due to the relative faintness of these objects for the {\it BRITE} satellites. 

One of the stars within the \brite PSF for \ec is  CPD\,$-$59$^\circ$2628 (V573 Carinae), which was discovered by \citet{2001A&A...369..561F} to be an eclipsing binary based on the long-term monitoring of \ec reported by \citet{1995A&A...304..415V} and \citet{1996A&AS..116....9S}. The period of the binary is 1.469332 d \citep{2001A&A...369..561F}, and the eclipse depths are nearly 0.5 mag in $y$-band photometry, and both the primary and secondary eclipses are similar in depth. We see evidence for this variability (at the half period, $f=$1.36 cycles d$^{-1}$) in our Fourier amplitude spectrum of \ec, which we marked with an arrow in the large panel on the right in Fig.~2.

\begin{figure}
\includegraphics[angle=90,width=3.3in]{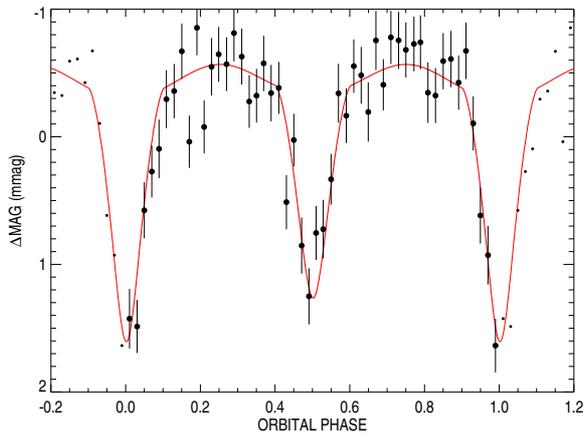}
\caption{The derived light curve of CPD\,$-$59$^\circ$2628 after heavily detrending the light curve of \ec, including a Wilson-Devinney model based upon the orbital parameters presented by \citet{2001A&A...369..561F}. The model was scaled to the observed variations to account for the heavy contamination from the surrounding stars. See text for further details.}
\label{eclipse}
\end{figure}

We were able to retrieve the light curve of the eclipsing binary through a careful de-trending of the light curve. Knowing the period of the binary (1.469 d) from the ephemeris reported by \citet{2001A&A...369..561F}, we did an averaging for each point across the orbital period of CPD\,$-$59$^\circ$2628. Then, we did a phase-averaging of the data in order to improve the detection of the eclipsing light curve. Fig.~\ref{eclipse} shows the data binned with $\sim$0.02 in phase. We also overplot a model of the system using a Wilson-Devinney code \citep{1971ApJ...166..605W} and the orbital parameters from \citet{2001A&A...369..561F} that was scaled to match the derived light curve from {\it BRITE}. The period derived from our data is of lower quality than that of \citet{2001A&A...369..561F}, but our time of minimum is HJD 2457707.9714$\pm$0.0040, with a phase offset of only $+0.0036\pm0.0040$ compared to the prediction of \citet{2001A&A...369..561F}. These results all show that the model and ephemeris of \citet{2001A&A...369..561F} remains accurate, and that our techniques yield high-quality results. 

Given the large number of nearby sources, we expect that the amplitudes of the pulsational signals found in the previous section will be lower than if the system were well-isolated and free of the Homunculus nebula. However, because of contamination the depth of the eclipse of CPD\,$-$59$^\circ$2628 is only $\sim0.5\%$ that of observed by \citet{2001A&A...369..561F}, which did not have the same level of contamination as these observations. Most likely, this system has the largest effect on the light curve of \ec. As such, the periodic photometric oscillations we detected with 2-mmag amplitudes would have to have an amplitude $>$0.5 mag if they originate in a field star instead of \ec. Among O-type stars, only eclipsing binaries are seen to have such large photometric amplitudes. 
Contamination can be an important issue with these \brite data. However, it also shows that the small satellites can provide useful information for some stars that are as faint as $V=10$.

\section{Discussion}

We have identified two independent frequencies listed in Table \ref{table} in the \ec system through high-precision photometric time series collected with {\it BRITE}. Additional frequencies, which are currently less than four-times the noise floor in significance, may become reliably detectable in subsequent high-precision photometric studies. 
The variability could also come from re-processed light from the primary in regions such as the Weigelt knots or wind collision zone. If the secondary contributes 2\% of the light at these wavelengths, which is a conservative upper limit from interferometric techniques, then the pulsations ascribed to these frequencies would have photometric amplitudes corresponding to 10\% of the star's luminosity. From the constraints on the secondary from analysis of the Weigelt knots surrounding \ec \citep{2010ApJ...710..729M}, the companion must be an early O star or WR star. As such a level of photometric variability from pulsations has not been observed in O stars \citep[e.g.,][]{PABLO, 2005ApJ...623L.145W} or WR stars \citep[e.g.,][]{2005ApJ...634L.109L, 2008ApJ...679L..45M}, it is reasonable to associate the oscillations with the primary star in the \ec system, and not the secondary star or the contaminating O stars mentioned in the previous section.

\begin{figure}
\includegraphics[angle=90,width=3.3in]{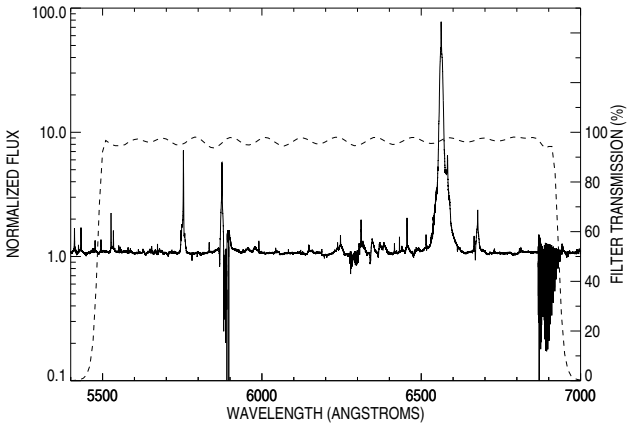}
\caption{A spectrum of \ec taken near apastron from \citet{2016MNRAS.455..244R} is shown with the \brite red filter transmission curve. The strong H$\alpha$ profile dominates this portion of the spectrum, which is why the y-axis is shown in logarithmic units. The contribution of H$\alpha$ is typically between 35--40\% of the light recorded by \brite, with additional contamination from the Homunculus nebula.}
\label{filter}
\end{figure}

\ec was previously studied photometrically by \citet{1995A&A...304..415V} and \citet{1996A&AS..116....9S}, who discovered a 58.58 d periodicity. This timescale is the same as our derived frequency $f_1$. \citet{1995A&A...304..415V} and \citet{1996A&AS..116....9S} found this oscillation to be relatively stable in frequency for over two decades of monitoring with Str\"omgren and Johnson photometry, and attributed it to pulsations. They did find that the amplitude of this pulsation was variable, caused by a decades-long beating effect with a 59.1 d period. Our data show that $f_1$ has an amplitude of 2.64 mmag, whereas the discovery of this period showed an amplitude of 35 mmag. However, there are some very important differences in these data. Our data include the entire Homunculus nebula \citep[as did the data in][]{1995A&A...304..415V, 1996A&AS..116....9S}, which can contribute a larger amount of flux in the $R-$ band than in the previous analyses; although, the exact contribution has not been measured precisely in recent years due to the continuing brightening of \ec making such a measurement impossible to do with {\it HST}. Further, the $R-$band is also contaminated by H$\alpha$, which contains emission from the primary's stellar wind and the wind-wind collision zone. Neither the contamination by the Homunculus or by H$\alpha$ can be easily corrected, but recent spectra by \citet{2016MNRAS.461.2540R} show that the H$\alpha$ contribution is at least $35-40\%$ of the observed light in the {\it BRITE} red filter near apastron, corresponding approximately to the orbital phase of the {\it BRITE} photometry reported here. We show a sample spectrum of the central region of the \ec system along with the \brite response filter in Fig.~\ref{filter}.

We suspect that the dilution of the pulsational amplitude by the neighboring stars, Homunculus, and the strong H$\alpha$ emission line may indicate that the pulsation has been stable for several decades. In fact, our derived times of maxima from the frequencies in Table \ref{table} are within a few days of the times of maxima reported in these previous studies. Given the similarity of our results to those of \citet{1996A&AS..116....9S} despite two different techniques being utilized, we should ask what the source of potentially coherent oscillations in the extreme primary star is. \citet{1996A&AS..116....9S} found that the primary pulsation period of 58.58 d showed a beat period of $\sim20$ years against a secondary period of $\sim59.1$d, which causes the observed photometric amplitude to vary. 
A secondary Fourier peak appears in the Fourier spectrum at a lower frequency than $f_1$, but would imply a beating period closer to 40 d after the removal of the frequencies $f_1$ and $f_2$. This frequency could also be emerging due to changes in the amplitude or phase of the signal. Future monitoring with {\it BRITE} may provide insights into the reality of the two similar frequencies and any beat frequencies in the data as our combined data set shows two emerging peaks in the region surrounding $f_1$ in the Fourier amplitude spectrum. A similar frequency spacing was resolved for \brite observations of HD 201433 through the addition of photometric data from another satellite \citep{2017arXiv170401151K}. We are hopeful that additional seasons of observations will elucidate these details in \ec. 

The only similar star that has shown stable pulsations is the LBV candidate MWC 314 \citep{2016MNRAS.455..244R}, which exhibits two pulsations with periods close to 1 d. This star has a spectrum reminiscent of the LBVs like $\eta$ Carinae, but many of the emission lines are double-peaked, forming in a circumbinary disk. In the case of MWC 314, \citet{2016MNRAS.455..244R} found that the binary was not a massive LBV as previously suggested by \citet{2008A&A...487..637M} and \citet{2013A&A...559A..16L}, but rather was the evolutionary byproduct of a lower-mass binary interaction. The system was a close spectroscopic twin to the binary HDE 326823 \citep{2011AJ....142..201R} where mass transfer has hidden the mass gainer behind a thick accretion torus while leaving the lower mass donor star visible. The analysis presented by \citet{2016MNRAS.455..244R} of MWC 314 showed that if the stars are rotating synchronously at periastron, then the mass of the primary is only 5--6 $M_\odot$. A second model, involving boosting the eccentricity of an orbit during mass transfer, showed the primary mass of MWC 314 was $\sim 14 M_\odot$. The pulsational behavior of MWC 314 that was observed with {\it MOST} confirmed the low-mass nature of this system \citep{2016MNRAS.455..244R}. Similar short-period ($\sim 6$ hr) pulsations were observed in the HD 5980 system by \citet{1997A&A...328..269S} after its LBV-like eruption, but the close orbit likely changes the observed stellar structure of these stars compared to other LBV stars. Similar coherent short-period time-scales are not seen in the light curve of \ec. 

Current models of \ec have a primary star weighing $\sim 90 M_\odot$ with a secondary mass of $\sim 30 M_\odot$ \citep{2012MNRAS.423.1623G}. These models have been used to self-consistently interpret high resolution, spatially resolved spectroscopy with {\it HST/STIS} \citep{2013MNRAS.436.3820M, 2016MNRAS.462.3196G}, the variability from the X-ray light curve \citep{2017ApJ...838...45C}, the variations from the colliding winds as observed with the He II $\lambda 4686$\AA\ emission line \citep{2016ApJ...819..131T}, and in variability from the He I P Cygni absorption troughs \citep{2016MNRAS.461.2540R}. The main constraints on the masses come from mass-luminosity relationships, though the luminosities are difficult to determine for \ec given the circumstellar environment \citep[e.g., ][]{2017ApJ...842...79M}. 

The two apparently coherent oscillations detected by {\it BRITE} could represent strange-mode pulsations that are expected to be the primary pulsations for stars that have gone through a red supergiant phase \citep{2013ASPC..479...47S}. Luminous blue variables have been observed to have non-periodic variations in their photometry lasting decades \citep{1994PASP..106.1025H} and these are thought to be driven by either an enhanced $\kappa$-mechanism or strange-mode oscillation which can account for many of the observed time-scales in LBVs and in normal supergiants \citep{2013ASPC..479...47S}. However, \ec represents one of the most luminous stars in the Galaxy and our understanding of stellar evolution suggests that such massive stars do not pass through the RSG phase.

\citet{1998A&A...335..605L} examined the variability of several LBVs and found that the microvariability may be explained by $g$-modes. They found periods ranging between 18 and 195 d, but all periods showed large changes over time scales of a few hundred days. Given the time sampling of the \brite photometry, $f_1$ may have been stable for many decades \citep{1995A&A...304..415V,1996A&AS..116....9S}. We also note that \citet{1999A&A...343..847V} found evidence of a 200 d timescale in the photometric behavior of $\eta$ Carinae, but our current \brite photometric dataset has too short a time-span to examine the behavior at that time scale.

However, neither strange-modes nor $g$-modes in LBVs have been shown to be stable in the Fourier domain over long timescales. Therefore, the most probable source of these oscillations may be tidally excited oscillations (TEOs), which seems plausible given our derived periods. While there has never been a system with such a long orbital period which has exhibited TEOs, \ec also has an unusually high eccentricity,
and masses. It is reasonable to suspect that these possible pulsations are TEOs as seen in several ``heartbeat" systems \citep[e.g.,][]{2017ApJ...834...59G}, even though they are observed far from the periastron event.

TEOs were first suggested by \citet{1941MNRAS.101..367C}, but not observed until recently \citep{2002MNRAS.333..262H, 2009A&A...508.1375M}. The analysis of TEOs became prominent with the advent of {\it Kepler} observations \citep[e.g., ][and references therein]{2016ApJ...829...34S}. Recent advances have provided a means to identify the geometry of the pulsational modes observed in these eccentric binaries \citep{2017ApJ...834...59G, 2014MNRAS.440.3036O}. In Table \ref{table}, we include potential harmonics to the 2022.7 d orbital period \citep[best derived recently by ][]{2016ApJ...819..131T} of $\eta$ Car. The identified modes all seem to have many potential orbital harmonics associated with them within the errors of the frequencies. The tidal driving of these modes is discussed from a theoretical standpoint by \citet{1995ApJ...449..294K}, with some of the clearest detections shown in the work by \citet{2012ApJ...753...86T}. 

The most massive confirmed heartbeat system is $\iota$ Ori \citep{PABLO}, which shows evidence for TEOs excited at 5 harmonics. The orbital harmonics at which these pulsations are observed are 23, 25, 27, 33, 75. In the case of $\eta$ Car, we would infer higher order harmonics, with the exception of $f_2$ in Table \ref{table}, which is near the 35th orbital harmonic of this system. The orbital period of \ec is also much longer than that of $\iota$ Ori (2022.7 d compared to 29.1 d), which, combined with the differing masses, might change the excited pulsational modes. If these are indeed TEOs, then \ec would exhibit both the longest known period for TEOs and would likely be the highest mass system to exhibit the phenomenon thus far.

Lastly, a massive binary in the Large Magellanic Cloud, R 81, contains a candidate LBV in a 78.5 d orbit that was characterized by \citet{2002A&A...389..931T}. The orbit is fairly eccentric ($e=0.569$), and shows photometric modulation around the orbit, with a possible eclipse in the system. \citet{2002A&A...389..931T} also found some relatively stable pulsations which have a period of 10.78 d, which are seen both photometrically and spectroscopically. The periods are a 7:1 harmonic of the orbital period, within the errors of both timescales measured. While the mass loss rate and terminal velocity are lower for R 81 than for $\eta$ Carinae, this may represent another massive system with a heartbeat and TEOs. This system would be the best analog for \ec if the pulsations discussed here are actually excited by the eccentric binary orbit. 

We tested our hypothesis of the tidally excited oscillations using the TIDES code \citep{1999RMxAA..35..157M,2011A&A...528A..48M}. The calculations assumed masses similar to those of most modern simulations of the system \citep{2013MNRAS.436.3820M}, namely the primary star had a mass of $100 M_\odot$ and a radius of $90 R_\odot$, with the secondary star having a mass of $30 M_\odot$. One parameter used in these simulations is the ratio of the rotation period of the primary star to the instantaneous circular orbital period at periastron. For these first calculations, we use a ratio of the two of 0.464, i.e., the star is rotating at 46\% the rate of the orbit at periastron. This synchronicity parameter was calculated from the orbital period, eccentricity, and an assumed radius ($R_p=100R_\odot$) and a rotational speed of 50 km s$^{-1}$. The uncertainty in the rotational speed is quite high, but some results indicate that the primary could be a rapid rotator \citep[e.g.,][]{2003ApJ...586..432S, 2010ApJ...716L.223G}. We want to further test models for the system once better constraints on the pulsational periods are derived from future {\it BRITE} data along with improvements on the orbital period from new measurements across the 2020 periastron passage. The results of the calculations are the radius as a function of orbital phase. We used this with the Stefan-Boltzmann law with $T_{\rm eff} = 27,500$ K to derive a light curve for the same orbital phase range as our {\it BRITE} observations. Our effective temperature is from the models of \citet{2001ApJ...553..837H} and represents the base of the wind, but effective temperature is not an easily defined quantity for this system.

From this simulated light curve, we extracted the portion from the same orbital phase range as these {\it BRITE} measurements. We de-trended the light curve in the same manner as the data reduction from the observations, and then calculated the Fourier amplitude spectrum of the simulated light curve. We scaled the Fourier transform to match the intensity of the strongest peaks in Figure \ref{FT}, shown as a blue line. The resulting Fourier analysis shows similar peaks as our derived $f_1$ and $f_2$ from {\it BRITE}. While the parameters of this specific simulation were not fine-tuned to our frequencies, they nevertheless show that interpreting the observed pulsations as TEOs is plausible, and that further observations with the {\it BRITE} nanosatellites will provide more precise pulsational frequencies, thus allowing us to fine-tune a model to derive fundamental parameters of the system. This interpretation provides a natural explanation for the long-lived nature of $f_1$ that was first reported from data collected more than two decades ago \citep{1995A&A...304..415V, 1996A&AS..116....9S}. Since our times of maxima seem to coincide with those of the previous studies, we suspect these modes to be tidally excited. We did not run a large grid of models to explore what frequencies and amplitudes are predicted for various values of rotation, radii, and masses. Our future analyses that incorporate more photometric data will provide a better dataset in which to model these pulsational frequencies.

We have begun planning observations to confirm these oscillations and refine the periods with \brite and to examine the possibility of tidally excited oscillations being the driving cause of this variability. We will also be working on obtaining high-quality spectroscopy of the system across the red filter from {\it BRITE}. This should confirm if the driving cause is from the continuum or the wind lines. Further, we will produce a refined set of models with the TIDES code \citep{1999RMxAA..35..157M,2011A&A...528A..48M} to constrain the fundamental parameters. With detailed modeling and observations, we will hopefully have a refined set of fundamental parameters. With accurate parameters from asteroseismology, we will begin to understand the evolution of the system.

\section*{Acknowledgements}

This work is based on data collected by the BRITE Constellation satellite mission, designed, built, launched, operated and supported by the Austrian Research Promotion Agency (FFG), the University of Vienna, the Technical University of Graz, the Canadian Space Agency (CSA), the University of Toronto Institute for Aerospace Studies (UTIAS), the Foundation for Polish Science \& Technology (FNiTP MNiSW), and National Science Centre (NCN).
The operation of the Polish BRITE satellites is funded by a SPUB grant by the Polish Ministry of Science and Higher Education (MNiSW).
NDR acknowledges postdoctoral support by the University of Toledo and by the Helen Luedtke Brooks Endowed Professorship, along with support of NASA grant \#78249. 
GK acknowledges support from CONACYT grant 252499.
APi acknowledges support from the NCN grant no.~2016/21/B/ST9/01126.
AFJM is grateful for financial aid from NSERC (Canada) and FQRNT (Quebec). 
AD is is supported by FAPESP Foundation.
GH acknowledges support by the Polish National Science Center (NCN), grant no. 2015/18/A/ST9/00578.
APo acknowledges NCN grant 2016/21/D/ST9/00656.
GAW acknowledges Discovery Grant support from NSERC.
WW was supported by the Austrian Research Promotion Agency (FFG).
KZ acknowledges support by the Austrian Fonds zur F\"orderung der wissenschaftlichen Forschung (FWF, project V431-NBL) and the Austrian Space Application Programme (ASAP) of the Austrian Research Promotion Agency (FFG).

\bibliographystyle{mnras}
\bibliography{etaBRITE}

\end{document}